\documentclass[conference,9pt]{IEEEtran}

\usepackage[svgnames,table]{xcolor}
\usepackage[T1]{fontenc}
\usepackage[tt=false, type1=true]{libertine}
\usepackage[scaled=0.8]{beramono}
\usepackage{microtype}

\usepackage{amsmath}
\usepackage{amsfonts}
\usepackage{relsize}
\usepackage{colortbl}
\usepackage{multirow}
\usepackage{array,makecell}
\usepackage{enumitem}
\setlist{topsep=1ex,itemsep=0ex,leftmargin=1em}
\usepackage{booktabs}
\usepackage{colortbl}
\definecolor{lightgray}{gray}{0.9}
\usepackage{url}
\usepackage{subcaption}
\usepackage{listings}
\usepackage{sra-color}
\usepackage{wasysym}
\usepackage{mdframed}
\usepackage{ marvosym }
\usepackage[all]{nowidow}

\usepackage{doi}
\usepackage{balance}

\usepackage{dataref}
\input{fig/data.dref}

\usepackage[backend=bibtex8,maxnames=8,maxcitenames=2,mincitenames=1,style=ieee]{biblatex}
\bibliography{paper}
\AtEveryBibitem{
  \clearfield{month}
  \clearname{editor}
  \clearlist{publisher}
  \clearlist{location} 
  \clearfield{venue}
}
\AtBeginBibliography{\small}

\usepackage{common}
\usepackage{xspace}
\usepackage{tikz}
\usetikzlibrary{tikzmark,calc,positioning}
\usetikzmarklibrary{listings}
\usepackage{etoolbox}
\usepackage[textsize=tiny,textwidth=1.4cm]{todonotes}
\definecolorseries{test}{rgb}{grad}[rgb]{.95,.55,.55}{11,11,17}
\resetcolorseries[10]{test}
\newcommand{\addtodoeditor}[1]{%
  \colorlet{#1}{test!!+!50}
  \expandafter\def\csname#1\endcsname##1{%
	\todo[color=#1,]{\sffamily\textbf{\uppercase{#1}:} ##1}\xspace%
  }
  \expandafter\newcommand\csname#1i\endcsname [1]{%
		\todo[inline, color=#1]{\sffamily\textbf{\uppercase{#1}:} {##1}}\xspace%
  }
}
\forcsvlist\addtodoeditor{tm,cd,mm}

\usepackage{csquotes}
\MakeOuterQuote"

\usepackage{tcolorbox}
\newtcolorbox{doibox}[1]{colback=safegreen!30,colframe=safegreen,fonttitle=\bfseries,title=#1}

\usepackage{acronym}

\acrodef{FI}{fault injection}
\acrodef{PUT}{program-under-test}
\acrodef{SEU}{single-event upset}
\acrodef{SDC}{silent-data corruption}
\acrodef{FS}{fault space}
\acrodef{SWIFI}{Software-Implemented \ac{FI}}
\acrodef{SAFI}{Simulation-Assisted \ac{FI}}
\acrodef{HAFI}{Hardware-Assisted \ac{FI}}
\acrodef{PC}{program counter}
\acrodef{ILP}{integer linear programming}
\acrodef{IPET}{implicit path enumeration technique}
\acrodef{WFFT}{linearly-weighted frequency spectrum}
\acrodef{DAG}{directed acyclic graph}

\usepackage{hyperref}
\usepackage{prettyref}
\newrefformat{cha}{\hyperref[#1]{Chapter~\ref*{#1}}}
\newrefformat{sec}{\hyperref[#1]{Sec.~\ref*{#1}}}
\newrefformat{sub}{\hyperref[#1]{Sec.~\ref*{#1}}}
\newrefformat{tab}{\hyperref[#1]{Tab.~\ref*{#1}}}
\newrefformat{fig}{\hyperref[#1]{Fig.~\ref*{#1}}}
\newrefformat{line}{\hyperref[#1]{l.~\ref*{#1}}}
\newrefformat{lst}{\hyperref[#1]{Lst.~\ref*{#1}}}
\newrefformat{pat}{\hyperref[#1]{Patch~\ref*{#1}}}
\newrefformat{alg}{\hyperref[#1]{Algorithm~\ref*{#1}}}
\newrefformat{eq}{\hyperref[#1]{Equation~\ref*{#1}}}
\newrefformat{app}{\hyperref[#1]{Appendix~\ref*{#1}}}

\usepackage{listings}
\usepackage{float}
\newfloat{listing}{thp}{lop}
\floatname{listing}{Listing}

\usepackage{siunitx}
\pgfkeys{
    /pgf/number format/.cd,
    set thousands separator={\,},
  }

\newcommand*{\eg}{e.g.,~}
\newcommand*{\ie}{i.e.,~}

\def\paragraph#1{\noindent\textbf{#1}\hspace{1em}}

\definecolor[named]{ACMPurple}{cmyk}{0.55,1,0,0.15}
\definecolor[named]{ACMDarkBlue}{cmyk}{1,0.58,0,0.21}
\hypersetup{colorlinks,pdfborder={0 0 0},
  linkcolor=ACMPurple,
  citecolor=ACMPurple,
  urlcolor=ACMDarkBlue,
  filecolor=ACMDarkBlue}
\def\orcid#1{\href{https://orcid.org/#1}{#1}}

\usepackage{soul}
\let\change\relax

\begin{document}

\title{Checkpoint Placement\\ for Systematic Fault-Injection Campaigns\\
{\footnotesize Accepted at ICCAD'23}}

\author{\IEEEauthorblockN{Christian Dietrich}
  \IEEEauthorblockA{\textit{Hamburg University of Technology}\\
    \orcid{0000-0001-9258-0513}\\
christian.dietrich@tuhh.de}
\and
\IEEEauthorblockN{Tim-Marek Thomas}
\IEEEauthorblockA{\textit{Leibniz Universität Hannover}\\
  \orcid{0009-0000-8197-2423}\\
  thomas@sra.uni-hannover.de}
\and
\IEEEauthorblockN{Matthias Mnich}
\IEEEauthorblockA{\textit{Hamburg University of Technology}\\
  \orcid{0000-0002-4721-5354}\\
  matthias.mnich@tuhh.de
}
}

\def\checkpoint{\textsc{Checkpoint}\xspace}
\def\knapsack{{\sc Knapsack}\xspace}
\def\uniform{\texttt{uniform()}\xspace}
\def\genetic{\texttt{genetic()}\xspace}
\def\ilp{\texttt{ilp()}\xspace}
\def\DP{\texttt{DP()}\xspace}

\maketitle
\drefkeys{PA/.style={assume math mode=true,precision=2,fixed,fixed zerofill}}
\begin{abstract}
Shrinking hardware structures and decreasing operating voltages lead to an increasing number of transient hardware faults, which thus become a core problem to consider for safety-critical systems.
Here, systematic fault injection (FI), where one program-under-test is systematically stressed with faults,  provides an in-depth resilience analysis in the presence of faults.
However, FI campaigns require many independent injection experiments and, combined, long run times, especially if we aim for a high coverage of the fault space.
One cost factor is the \emph{forwarding phase}, which is the time required to bring the system-under test into the fault-free state at injection time.
One common technique to speed up the forwarding are checkpoints of the fault-free system state at fixed points in time.

In this paper, we show that the placement of checkpoints has a significant influence on the required forwarding cycles, especially if we place faults non-uniformly on the time axis.
For this, we discuss the checkpoint-selection problem in general, formalize it as a maximum-weight reward path problem in graphs, propose an ILP formulation and a dynamic programming algorithm that find the optimal solution, and provide a heuristic checkpoint-selection method based on a genetic algorithm.
\change{Applied to the MiBench benchmark suite}, our approach consistently reduces the forward-phase cycles by at least \dref[PA,percent,precision=0]{/by-caches/benchmark/genetic/min/value} percent and up to \dref[PA,percent,precision=3]{/by-caches/benchmark/genetic/max/value} percent when placing 16 checkpoints.
\end{abstract}


\section{Introduction}

Functional safety standards (\eg ISO 26262 or IEC 61508 \cite{ISO26262-9,IEC61508}) demand that we assess the effects of transient hardware faults (soft errors) on our systems.
As soft errors are rare in reality~\cite{sridharan:13:sc,li:10:atc}, it is common to use systematic \ac{FI}~\cite{arlat:90:ieeese,ziade:04:ajit} to quantify the resilience of a program.  
Such systematic \ac{FI} campaigns are typically executed in three steps:
\begin{figure}[h]
  \includegraphics[width=\linewidth]{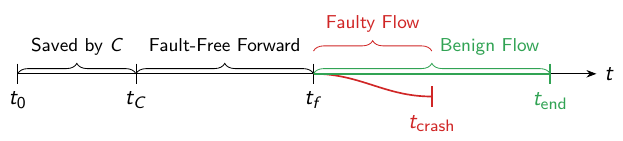}
  \caption{Phases of an Injection Experiment: The golden run spans $[\tStart, \tEnd]$, the injection is done at \tFI, and the checkpoint $C$ restores the program state at \T{C}}
  \label{fig:model}
\end{figure}

\begin{figure}[t]
  \centering
  \includegraphics[page=1,width=\linewidth]{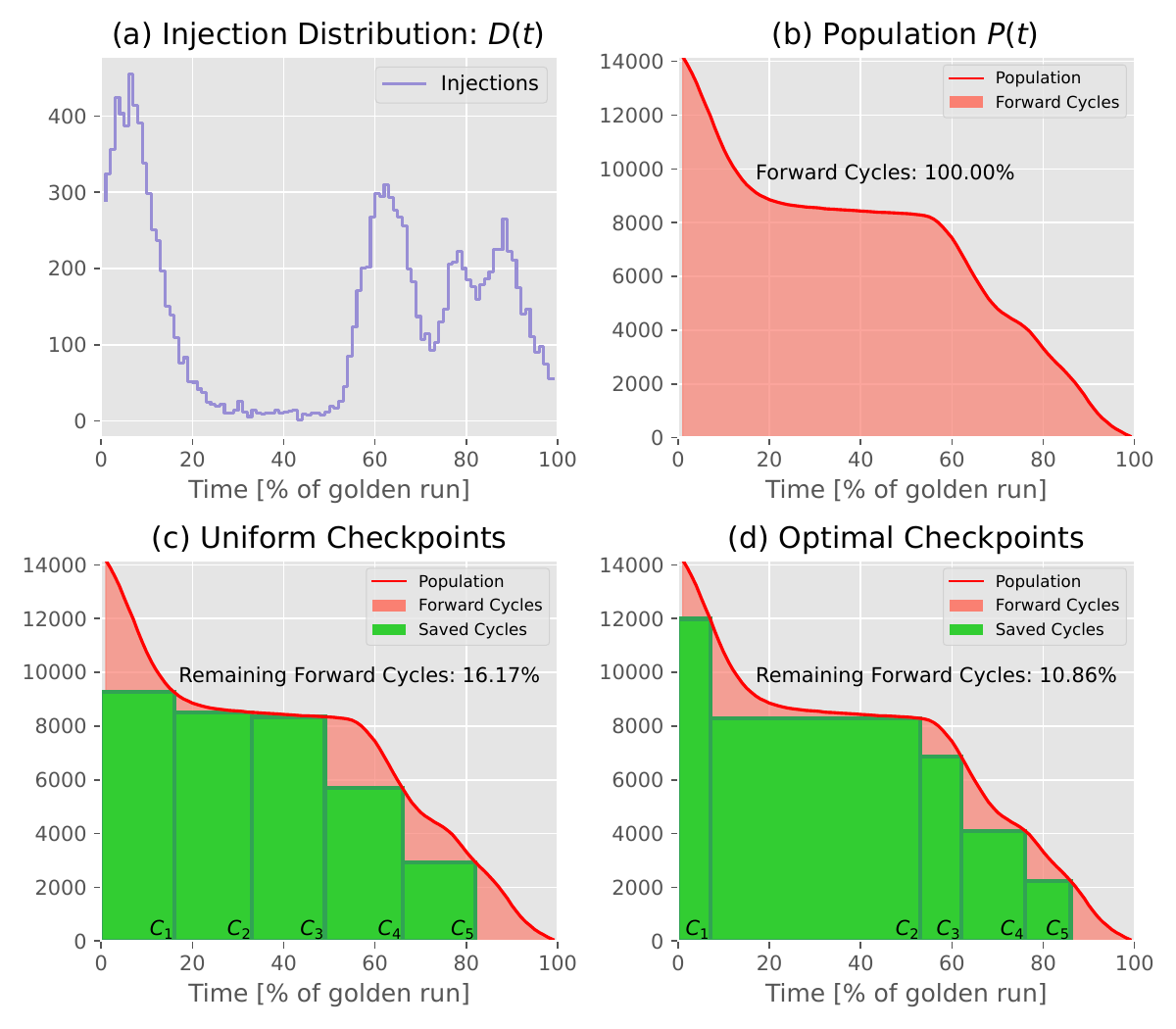}
	\caption{Checkpoints on an example FI distribution. (a) The distribution of the 14,000 injection experiments determines the (b) population of the required per-experiment forward cycles. Its integral (red area) is the total number of forward cycles for the FI campaign, which can be reduced by checkpoints (green areas). The (c) uniform placement of checkpoints is the state of the art, which is significantly outperformed by (d) an optimal checkpoint-selection algorithm.}
  \label{fig:forward-cycles}
\end{figure}

\begin{enumerate}
\item[(1)] \emph{trace} a fault-free program execution as the \emph{golden run}, which spans up the \ac{FS} of all potential faults (one fault per every time step and for every bit of information).
\item[(2)] \emph{prune} the \ac{FS}~\cite{smith:95:rms,guthoff:95:ftcs,pusz:21:lctes} to plan a representative subset of faults as \emph{pilot injections}, which will be carried out.
\item[(3)] re-execute the program for every planned pilot, \emph{inject} it at the planned time and location, and classify the following program behavior.
\end{enumerate}
For each injected fault $f$ in step (3), we need to bring the \ac{FI} platform into the fault-free state at \tFI (\prettyref{fig:model}):
After a reset to \tStart, the platform \emph{forwards} the program by fault-free execution to \tFI.
There, we inject~$f$ and continue the---now faulty---execution flow until the \ac{FI} platform detects a crash, a completion, or a timeout:

\noindent
Depending on the \change{program-under-test, the} fault model, and the employed pruning strategy in step (3), the distribution $D(t)$ of pilot injections over the run time $t$ is typically \emph{not} uniform.
For instance, if DRAM is unprotected while processor caches employ ECC, potential soft errors manifest whenever \change{the program loads} new data into the cache. 
A pruning strategy that takes this into account would yield a \ac{FI} distribution as shown in \prettyref{fig:forward-cycles}~(a).
\change{Please note that selecting pilot injections, is \emph{not} subject of this paper. For our proposed method, the distribution $D(t)$ of planned/executed fault injections is given.}

The later the fault, the more forward instructions are executed before the actual injection takes place; earlier instructions are forwarded more often than later instructions (\prettyref{fig:forward-cycles}~(b)).
For this, we introduce the term \emph{population}, which is the number of experiments that are in the forward phase at a given point in time if we would start them all in parallel at \tStart.
With \tFI of a fault $f$, that fault leaves the population; making it a monotonically decreasing function.

A broadly applied technique to speed up the (repetitive) forwarding is  \emph{checkpointing}~\cite{king:05:atc,berrojo:02:date}:
By resetting the system not to the initial state at~\tStart, but to some later state at $\T{C} \le \tFI$ (\prettyref{fig:model}), we save a significant amount of forwarding cycles.

In this paper, we address the question of \emph{checkpoint placement}.
The state-of-the art approach~\cite{hari:14:isca,berrojo:02:vlsi,berrojo:02:date,tuzov:16:ladc,rosa:15:dfts} is to distribute checkpoints \emph{uniformly} on the time axis (\prettyref{fig:forward-cycles}~(c)).
However, this is not ideal: Significantly higher savings can be obtained by our optimal placement strategy of the checkpoints (\prettyref{fig:forward-cycles}~(d)).
For the paper, we claim the following contributions:

\begin{itemize}
\item We describe the \checkpoint problem for a fixed number of checkpoints and reduce it to a maximum-weighted reward path problem for \acp{DAG}.
\item We point out the shortcomings of the naive time-uniform checkpoint selection and propose three distribution-dependent checkpoint-selection methods.
\item We quantify the benefits of our approach on real-world \ac{FI} distributions and show that our methods outperform the uniform selection in the best case by \dref[percent]{/by-caches/benchmark/delta/max/value} percentage points.
\end{itemize}

The remainder of this paper is organized as follows:
\prettyref{sec:system-model} describes our \ac{FI} model and characterizes the checkpoint-selection problem.
In \prettyref{sec:method}, we reduce the problem to a constant-length maximum-weight reward path in a transitive \ac{DAG} and provide three selection methods.
We evaluate and compare all methods in \prettyref{sec:evaluation}, discuss our findings in \prettyref{sec:discussion}, review  the relevant literature in \prettyref{sec:related-work}, and conclude the paper in \prettyref{sec:conclusion}.

\section{Fault-Injection Model and Problem Description}
\label{sec:system-model}

\paragraph{Systematic FI Campaigns}
We target \emph{systematic} \ac{FI} campaigns, which plan and inject many different faults into deterministic re-runs of the same \ac{PUT}.
From a fault-free golden run, which has the temporal extent $[\tStart, \tEnd]$, a fault-planning strategy chooses $F$ faults for injection.
For each fault $f$,
we bring the \ac{FI} platform (\eg simulator or FPGA) into the fault-free state at \tFI (see \prettyref{fig:model}):
After a reset to \tStart, the platform \emph{forwards} the program by fault-free execution to \tFI.
There, we inject $f$ and continue the, now faulty, execution flow until the \ac{FI} platform detects a crash, a completion, or aborts the injection (\eg timeout).

With checkpoints, we cut the fault-free forwarding time:
Instead of resetting to \tStart, we restore to a previously-saved checkpoint $C$, which brings us directly into the fault-free state at \T{C}.
From there, we endure a shorter forwarding phase, which saves us $\T{C}-\tStart$ cycles for \emph{this} injection (see \prettyref{fig:model}).
A checkpoint \C{i} is usable for all faults with $\T{\C{i}} \le \tFI$, but it should only be used for faults with $\T{\C{i}}
  \le \tFI < \T{\C{i+1}}$ to maximize savings.

We assume that we can perform exactly $k$ checkpoints at different points in time.
This is, for example, the case when we employ FPGAs as \ac{FI} platform \cite{nowosielski:15:date} and use duplicated flip-flops to store the checkpoint, which allows for checkpoint restoration in a single cycle but limits $k$ to the number of FPGAs.
We further assume that checkpoint restoration is equally fast or faster than a full reset, which is inherently the case if resets are also implemented by a checkpoint at \tStart (as in GemFI~\cite{parasyris:14:dsn} or MEFISTO~\cite{jenn:94:ftcs}).
We further assume that the time axis is discretized into equal intervals, for whose we use the term \emph{cycle}.

\paragraph{Checkpoint Selection Problem}
Our goal is to reduce the number of required forwarding cycles over all planned fault injections $F$ with $k$ checkpoints.
To give an intuition for this problem, we look at the relation between fault distribution, checkpoints, and the number of forwarding cycles in \prettyref{fig:forward-cycles}:
In \prettyref{fig:forward-cycles}~(a), we show an artificial \ac{FI} distribution $D(t)$
over $[\tStart, \tEnd]$, where over 400 injections happen around $t=10$ while almost none are at $t=40$.

\prettyref{fig:forward-cycles}~(b) shows the \ac{FI}-experiment \emph{population} $P(t)$ that execute a specific forward cycle if we do \emph{not} employ checkpoints:
At $t=0$, all \num{14000} experiments start and run until their respective $\tFI$, where they leave the forwarding population, whereby $P(t)$ is a monotonically decreasing function from $P(\tStart)$ to $P(\tEnd)=0$.
Formally, the non-checkpointed \ac{FI} population is

\[
  P(t) = \int_{t}^{\tEnd} D(t)
\]

As each running experiment executes $\tFI-\tStart$ forwarding cycles, the sum of all forwarding cycles for the whole \ac{FI} campaign is equal to the integral $\int_{\tStart}^{\tEnd} P(t)$.
It is our overall goal to shrink this integral.

In \prettyref{fig:forward-cycles}~(c) and (d), we see how checkpoints achieve this goal:
At $\T{\C{i}}$, only those faults $f$ enter the population whose $\tFI \in [\T{\C{i}}, \T{\C{i+1}})$.
Each checkpoint \enquote{cuts out} a (green) rectangle of area $w$ and we end up with the vastly reduced red areas (see (c) and (d)), which however vary depending on the position of the checkpoints.
More formally, the set of checkpoints \Cset, whose size is determined by the \ac{FI} platform, saves us $S(\Cset)$ forwarding cycles over all fault injections.
For notational ease, we use the system reset at \tStart as an artificial checkpoint \C{0}.

\[
  S(\Cset) = \sum_{i=0}^{k-1} w^{\C{i}}_{\C{i+1}} = \sum_{i=0}^{k-1} (\T{\C{i+1}}-\T{\C{i}}) \cdot P(\T{\C{i+1}})
\]

\noindent After these fundamental considerations, we can describe the \checkpoint problem more precisely:
Given $P$ and $k$, where should we place our checkpoints to maximize $S(\Cset)$?

\section{Checkpoint Selection}
\label{sec:method}

\begin{table}[b]
  \centering
  \begin{tabular}{lp{5cm}}\toprule
    Variables & Description\\\midrule
    $t_0\ldots t\ldots \tEnd$  & discrete time axis\\
    $D: t \mapsto \mathbb{N}_0$ & \ac{FI} distribution \\
    $P: t \mapsto \mathbb{N}_0$ & \ac{FI} population, integral of $D$. \\
    $s_0\ldots s_n$ & steps in $P()$, there are $n$ steps. \\
    $\{C_1\ldots C_k\} \in \Cset$ & set of $k$ (real) checkpoints\\
    $C_0 = s_0 = t_0$ & artificial checkpoint to model system reset\\
    $R^a_b$ & the rectangle below $P$ from $(s_a,0)$ to $(s_b, P(s_b))$ and with area $w_{a,b}$.\\
    \bottomrule
  \end{tabular}
  \caption{Notation Overview}
  \label{tab:notation}
\end{table}

The state-of-the-art checkpoint selection strategy, termed \uniform, involves evenly distributing checkpoints along the time axis between~\tStart and \tEnd.
While this method is frequently employed in the literature~\cite{hari:14:isca,berrojo:02:vlsi,berrojo:02:date,tuzov:16:ladc,rosa:15:dfts}, it neglects the \ac{FI} distribution and fails to utilize a degree of freedom available on the time axis.
Consequently, it is evident that \uniform typically does not maximize savings.
However, given its negligible computation cost, $S(\Cset)$ can directly contribute to end-to-end savings without the need to offset any selection overheads.
Therefore, any checkpoint-selection methodology has to strike a balance between the production of optimal results and computational overhead.

Moving forward, we will delve into the underlying computational complexity of the checkpoint-selection problem and introduce three distinct solution strategies:
\ilp, \DP, and \genetic.
\prettyref{tab:notation} gives an overview of our notation.

\subsection{Theoretical Considerations}
\label{sec:reduction}
First, we aim to examine the structure of the \checkpoint problem to gain insight into its computational complexity.
Despite the qualitative nature of \prettyref{fig:forward-cycles}\,(b), $P(t)$ is, in reality, a step function with a discrete abscissa (time axis) and steps $s_0\ldots s_n$.
The maximum number of steps is $n \le \tEnd$, but fewer steps may be present if no experiments are scheduled for a certain $t$  ($D(t) = 0$).
However, the number of steps is generally large.

Moreover, we only need to consider these steps as potential locations for checkpoints, because moving a checkpoint $c_j$ that is located between two steps ($s_i < c_j< s_{i+1}$) to $s_{i+1}$ will always increase savings by $(s_{i+1} - c_j) \cdot P(s_{i+1})$.
Therefore, we can translate any \Cset to a $\Cset'$ with $S(\Cset') \ge S(\Cset)$ by relocating all checkpoints to the next step.

We associate \checkpoint with finding the maximum-weight reward path of constant length in a \ac{DAG}.
This problem was recently connected~\cite{axiotis:18:arxiv} to the \knapsack problem, a foundational problem in combinatorial optimization, known to be $\mathsf{NP}$-complete~\cite{karp:72:ccc}.
However, in a graph with $m$ edges, we can find the maximum-weight reward path of length $k$ using dynamic programming in $\mathcal{O}(km)$ time~\cite{axiotis:18:arxiv}.
Yet, we are not aware of any implementation in an actual application of that theoretical result.

The correlation with \checkpoint is as follows:
consider our population function $P$, with steps $s_0,\hdots,s_n$ at discrete time steps on the abscissa, where $s_0 < \hdots < s_n$.
For notational simplicity, we define $s_0 := 0$.
In our reduction, we create several rectangles for each time step $s_t$; the height of each rectangle is chosen to be~$P(s_t)$, fitting precisely beneath the curve $P$ at $s_t$.
For each step~$s_t$, we create exactly $t$ rectangles $R_t^0,\hdots,R_t^{t-1}$ and set the width of $R_t^i$ to $s_t-s_i$ for $i = 0,\hdots,n$.
The rectangle's area $w$ equals its height times its width.
The total number of rectangles created by this reduction is $\sum_{i=0}^{n} i = \mathcal{O}(n^2)$.
The optimal checkpoint selection is then equivalent to find $k$ rectangles that maximize the covered area under the $P()$.

To construct a \ac{DAG}, we create one node $v_t$ for each step $s_t$ and introduce the artificial entry and exit nodes ($v_0$, $v_n$) that act as start and end points for our desired paths.
Moreover, we add one arc $e_{i,j}=(v_i, v_j)$ for each rectangle $R^i_j$ we created and set its weight to the rectangle area $w_{i,j}$.
Our graph is directed (\ie in a positive direction), acyclic (\ie no backward edges), and complete (\ie transitive).
Finding $k$ optimal checkpoints can then be stated as finding the maximum-weight path between $v_0$ and $v_n$ that visits~$k$ inner nodes.

The intuition behind this reduction is that selecting $k$ checkpoints is, in fact, a step-wise under-approximation of $P()$ with $k$ rectangles.
If maximized, this approximation minimizes the integral of the error to $P()$; this integral comprises the remaining forwarding cycles of our \ac{FI} campaign.
In our reduction, we created all possible rectangles under $P()$ that span between two steps.
But to calculate a $k$-stepped under-approximation of $P$, we must (a) avoid selecting overlapping rectangles, (b) ensure no gaps between adjacent rectangles, and (c) select $k$ rectangles so their combined width is $\tEnd-t_0$.

Our graph structure encodes these constraints:
(a) if we select the arc $e_{i,j}$ we cannot select an arc $e_{a,b}$ with $a \le j$ and $b \ge i$ because the graph contains no backward arcs.
(b) If we enter an inner node~$v_i$ via the arc~$e_{\ell,i}$, we also must visit a leaving arc~$e_{i,r}$ to reach the non-inner node $v_n$, meaning all selected rectangles "touch" each other.
(c) As the arc $e_{i,j}$ represents the rectangle $R^i_j$ with width $s_j-s_i$, and we select only adjacent rectangles between~$v_0$ and $v_n$, our selection spans from $t_0$ to \tEnd.

\subsection{ILP-Based Checkpoint Selection}\label{sec:ilp}
\begin{figure}[t]
  \centering
  \includegraphics[page=1,width=0.9\linewidth]{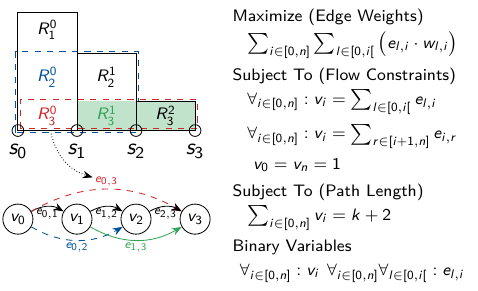}
  \caption{ILP formulation of the checkpoint selection problem.}
  \label{fig:ilp}
\end{figure}

To operationalize our reduction of \checkpoint, we use \emph{integer linear programming} (ILP)\acused{ILP} to formulate the selection of checkpoints as the optimization problem \ilp.
More precisely, we use the \ac{IPET}, which is also widely used in the real-time domain~\cite{puschner:1997:rts,li:1995:sigplan} to find the worst-case execution path through a program.
In contrast to control-flow graphs, our graph is acyclic, arc-weighted, and we search for a constant-length path.

For our \change{IPET} formulation (see \prettyref{fig:ilp}), we introduce one binary variable~$v_t$ for every inner node in the previously described DAG. \change{If the ILP solver sets $v_t$ to one, this indicates that step $s_t$ is selected as a checkpoint.}
Further, we introduce the artificial nodes $v_0$ and $v_n$, which act as entry and exit nodes to our \ac{DAG}.
In our example, $v_3$ is $v_n$.
With the constraint $\sum v_t = k+2$, we force the solver to select a constant-length path through our \ac{DAG}, visiting exactly $k$ inner nodes; placing $k$ \enquote{real} checkpoints.

With \change{IPET} flow constraints, we encode the \ac{DAG} structure:
For each arc, we introduce a binary variable $e_{i,j}$ that determines whether the arc from $v_i$ to $v_j$ is on the chosen path.
For each inner node, the sum of incoming ($e_{*,t}$) and the sum of outgoing ($e_{t,*}$) must be equal to the node variable $v_t$.
The intuition behind this is that an inner node is entered as often as it is left.
Further, the entry and exit nodes are surely part of the chosen path.
As our graph has no cycles, each node and each arc can be visited exactly once and all variables have a binary domain. 

As maximization objective, we sum up all arc variables, which we weight by the area $w$ of the rectangles they represent.
For example, $e_{0,2}$ is weighted with $w_{0,2}=(s_2 - s_0) \cdot P(s_2)$ (blue dashed rectangle).
By construction, our ILP formulation will result in the optimal checkpoint placement for a given \ac{FI} distribution.
In total, we require $\mathcal{O}(n^2)$ many binary variables for a distribution with $n$ steps.
For our example, with $k=1$, the solver has only one degree of freedom in choosing an inner node for a checkpoint.
The possible solutions are $\{v_0, v_1, v_3\}$ and $\{v_0, v_2, v_3\}$, which reflect the paths $e_{0,1} \rightarrow e_{1,3}$ and $e_{0,2} \rightarrow e_{2,3}$.

\subsection{Checkpoint Selection with Dynamic Programming}

As solving integer linear programs is usually computationally expensive, and moreover does not come with any worst-case run times in general, we further provide a dynamic programming algorithm \DP to find the maximum-weight reward path in an arc-weighted DAG $G$. 
We use $v_0$ and $v_n$ as DAG entry and exit nodes (source and sink), and search for a path with length $k+1$ and maximum weight between $v_0$ and $v_n$.

We create a dynamic programming table $T$, which contains entries~$T[i,j]$ which encode the \emph{maximum} weight of any path in $G$ which starts at $v_0$ and ends at $v_i$, uses at most $j$ internal nodes from $v_1$,\ldots,$v_i$. 
Thus, the table has $n \cdot (k+1)$ entries.

We initialize the table by setting $T[i,0] = w_{0,i}$ for $i \in [0, n]$ where and $w_{a,b}$ is the weight of the edge between node $v_a$ and~$v_b$; further, $w_{a,a} = 0$ for all nodes $a$.
We compute all other entries $T[i,j]$ with $j > 0$ recursively through
\[
  T[i,j] = \max_{x=0}^i \left\{ T[i-x, j-1] + w_{i-x, i}\right\} \enspace .
\]
The correctness of this recursion follows from the fact that to compute the $j$-step maximum-weight path between $v_0$ and $v_i$, we consider all possibilities for the additional step being located at $v_{i-x}$.
Left of $v_{i-x}$, we have a $j-1$-step path of weight $T[i-x,j-1]$, while we append one additional step of weight $w_{i-x, i}$ to the right of $v_{i-x}$.
We fill the table $T$ step by step for increasing values of $i$ and $j$ and read off the maximum weight of a solution in the table entry $T[n,k]$.
All entries are positive, and for each step only values from the previous row $j-1$ are required.

To identify the inner nodes (\ie the selected checkpoints) that are part of the maximum-weight path, we use a second table $X$ of size $n\cdot (k+1)$ that records the value $x$ for the selected maximum as $X[i,j] = x$.
Afterwards, we set $C_j = X[C_j+1, j]$ with $C_k=X[n, k]$.

As we have to consider up to $n$ possibilities for each entry and there are $n \cdot (k+1)$ entries, the whole procedure takes $\mathcal{O}(k \cdot n^2)$ time and $\mathcal{O}(k \cdot n)$ space.
In the checkpoint-selection case, it is likely that we can further reduce the computation complexity by using that our DAG is a complete graph (\ie a tournament), and moreover, the arc weights may satisfy the triangle inequality.


\subsection{Genetic Checkpoint Selection}
\label{sec:genetic-algorithm}

As the computation time requirement of \DP is still quadratic, we propose the heuristic checkpoint-selection strategy \genetic based on genetic algorithms~\cite{holland:73:siam}.
\change{We have chosen genetic algorithms as (1) checkpoint selection is a discrete optimization problem and (2) we expect that combining two good solutions (cross-over operation) will often yield an even better result.}
Using a heuristic also allows us to abort the selection process when we see no further progress or when a pre-defined time budget runs out.
We describe \genetic by defining the used genome, the genome-derived phenotype, the fitness function, as well as the used mutation operators.

For the genome, which encodes one valid checkpoint selection, we choose an $k$-sized vector of steps from $P$.
As phenotype, we use the rectangles spanned by the selected steps and use $S(\Cset)$ as the fitness function, which we aim to maximize.

As \emph{random} mutation operators, we combine two genomes by a two-point crossover with p=0.5, or mutate the one checkpoint by (each with p=0.125):
(1) moving it one step to the left/right,
(2) moving it three steps to the left/right,
(3) moving it to a random step,
or moving it to the middle between its left and right neighbor.

Initially, we start with a population of 100 random genomes.
In each round, we enlarge this population by cross-over and mutation to 300 individuals.
After sorting them according to the fitness function, we surely select the 10 best genomes and exchange place 11 to 100 with p=0.5 with another randomly-picked individual to avoid getting stuck in a local optimum.
We execute this heuristic search up to a given number of seconds in parallel and return the globally-best selection of checkpoints.
The described algorithm does \emph{not} guarantee an optimal solution.


\section{Evaluation}
\label{sec:evaluation}

With our evaluation, we demonstrate that %
(1) \uniform shows a wide range of forward-cycle reductions,
(2) our selection methods produce consistently better (or equal) results than \uniform,
(3) the achieved advantage correlates with the non-uniformity of the fault distribution, and
(4) \genetic's results were optimal for multiple hundreds of distributions but at lower costs than \ilp and \DP.

We compare our methods by saved cycles, runtime, and sensitivity to the uniformity of the distribution. We consider \uniform to be the state-of-art checkpoint-selection method (see \prettyref{sec:related-work}).
For our evaluation, we use synthetic benchmarks and realistic \ac{FI} distributions, which we derive from the MiBench benchmark suite~\cite{guthaus:01:wwc}.

\subsection{Non-Uniformity Metric}

As we already discussed, the shortcoming of \uniform is that it does not exploit the temporal variance of the \ac{FI} distribution.
To quantify the intuition that \genetic can perform better on less uniform distributions, we require an metric to measure the \enquote{non-uniformity}~$U^-$ of the distribution $D$.
For this, we normalize it as $\overline{D}$ in time and height to 100 percent and propose the \ac{WFFT} metric with $\text{fft}_i$ being the $i$-th element of the fast Fourier transformation:

\[
  U^-(\overline{D}) = \sum_{i=0}^{100} i \cdot \left|\text{fft}_i(\overline{D})\right|
\]

With this metric, we look at the distribution in the frequency domain and weight low-frequency signal shares with a higher value than high-frequency shares.
Thereby, the perfectly-uniform distribution will result in $U^-=0$ and distributions with larger gaps end up with a higher score.

\subsection{Synthetic FI Distributions}

\begin{figure}
  \centering
  \includegraphics[width=\linewidth]{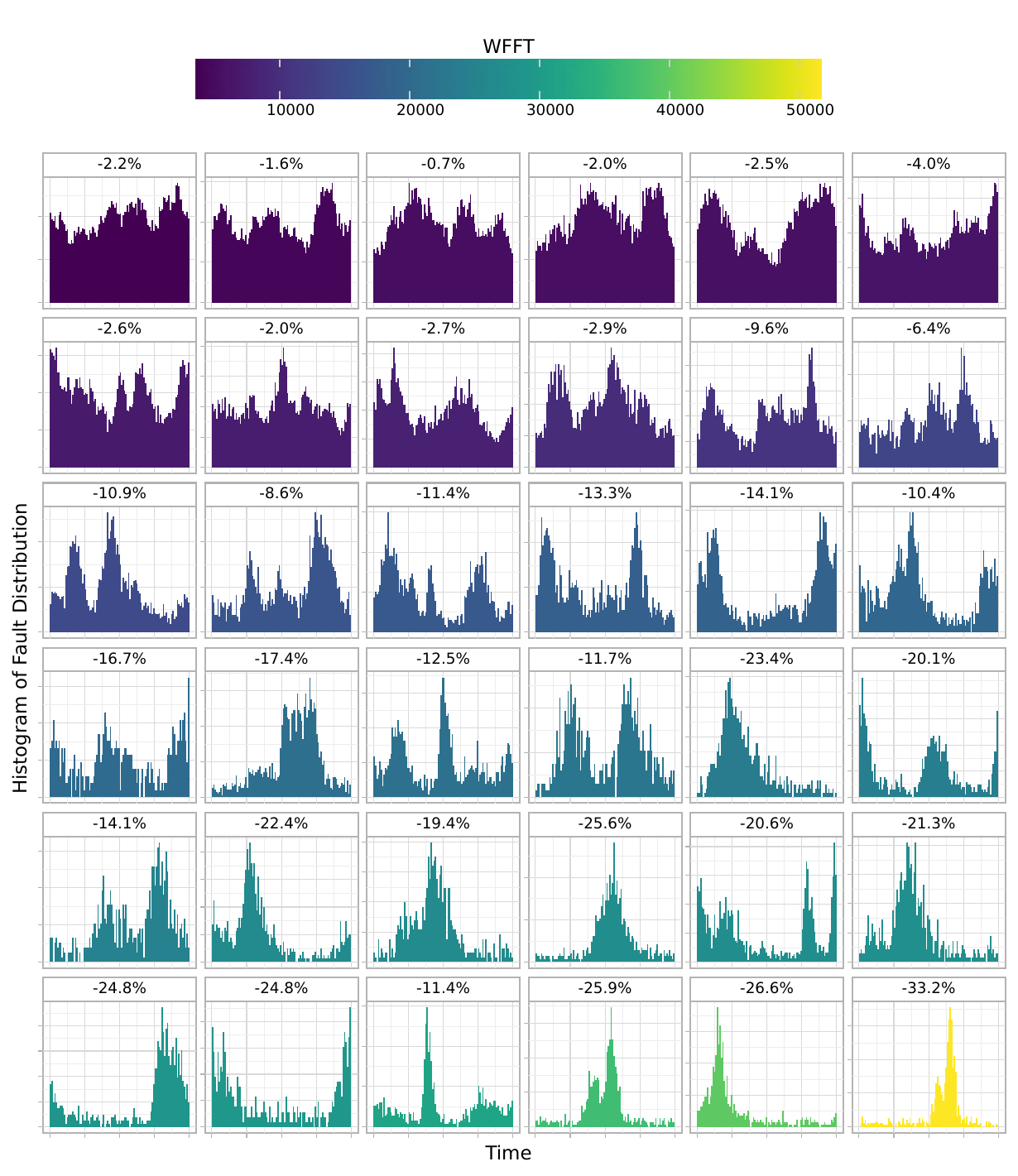}
  \caption{
    \change{36} randomly generated distributions, colored and sorted by their \ac{WFFT} \change{and arranged in left-to-right and top-to-bottom order}. Each tile is annotated with the percentual forward-cycle reduction that \genetic achieves over \uniform for placing 8 checkpoints.
  }
  \label{fig:random_wfft}
\end{figure}

Our initial objective is to compare the \uniform, \genetic, and \ilp approaches on synthetic \ac{FI} distributions.
These synthetic distributions are designed to qualitatively emulate \change{different} real-world distributions (see also \prettyref{fig:distr}), while offering the flexibility to span a wide variety of distributions and degrees of non-uniformity.
The use of synthetic benchmarks enables us to scale \tEnd, which in turn determines the number of steps $s_t$ and consequently the number of ILP variables.
\change{As different programs and/or pruning methods only differ in the fault distribution at the checkpoint-selection stage, our evaluation demonstrates the generalizability of our approach.}

\paragraph{Generation of \ac{FI} Distributions}
To synthesize random distributions, we start with a uniform distribution of faults which serves as a "noise carpet".
Onto this carpet, we overlay between~2 and 100 (log-normal distribution) peaks shaped by the Gumbel distribution~\cite{gumbel:41:ams}.
Each peak's height ranges from 2 to~5 times (uniform distribution) that of the carpet, with a width constituting 2 to 10 percent (uniform) of the total distribution.
The Gumbel peaks simulate localized FI maxima, while the uniform carpet establishes a "base height".

To illustrate the end results of our distribution generation method, refer to \prettyref{fig:random_wfft} which presents 36 FI histograms with 10000 steps, sorted and color-coded by their \ac{WFFT}.
Our generation approach produces distributions with diverse characteristics and a broad spectrum of non-uniformity.
For instance, the top row displays more uniform distributions punctuated by a few shallow peaks, indicative of a low WFFT.
In contrast, the bottom row exhibits distributions with a high WFFT, resulting from a few pronounced peaks and a smaller portion of uniformly distributed faults.

Each tile in \prettyref{fig:random_wfft} is additionally annotated with the percentual savings in forward cycle achieved by the \genetic algorithm over the \uniform method when \emph{eight} checkpoints are placed.
The additional savings range from \dref[precision=1,fixed]{/random/genetic_improvement/min/value} to \dref[precision=1,fixed]{/random/genetic_improvement/max/value} percent.
It is noteworthy that less uniform distributions, as denoted by a higher WFFT, tend to yield larger \genetic gains.
This trend is especially pronounced in distributions characterized by a few high peaks, as the \genetic algorithm can effectively pinpoint these to position the checkpoints, unlike the \uniform method which remains oblivious to the distribution characteristics.

\begin{table}[t]
  \centering
  \drefkeys{prefix=/random/ilp,precision=1,fixed,fixed zerofill,assume math mode=true}
  \def\row#1{#1
    & \dref{/#1/ilp_mean}$\pm$\dref{/#1/ilp_std}\,s
    & \dref[scale by=1000]{/#1/dp_mean}$\pm$\dref[scale by=1000]{/#1/dp_std}\,ms
    & \dref{/#1/genetic_mean}$\pm$\dref{/#1/genetic_std}\,s
    & \dref[precision=0]{/#1/genetic_optimal}/\dref[precision=0]{/#1/count}
  }
  \def\rowBig#1{#1
    & --
    & \dref{/#1/dp_mean}$\pm$\dref{/#1/dp_std}\,s
    & \dref{/#1/genetic_mean}$\pm$\dref{/#1/genetic_std}\,s
    & \dref[precision=0]{/#1/genetic_optimal}/\dref[precision=0]{/#1/count}
  }
  \caption{Scalability on Synthetic Distributions}
  \begin{tabular}{lrrrr}\toprule
          & ILP       & DP & \multicolumn{2}{c}{Genetic Algorithm}\\\cmidrule(r){2-2}\cmidrule(r){3-3}\cmidrule(r){4-5}
    Steps &  Run Time & Run Time & Run Time & Optimal? \\\midrule
    \row{500}\\
    \row{1000}\\
    \row{1500}\\
    \row{2000}\\
    \row{2500}\\
    \row{3000}\\\midrule
    \rowBig{10000}\\
    \rowBig{50000}\\
    \rowBig{100000}\\
    \rowBig{150000}\\
    \bottomrule
  \end{tabular}
  \label{tab:ilp}
\end{table}

\paragraph{Scalability}
With our synthetic distributions, we want to compare the scalability of \ilp, \DP, and \genetic.
This demonstrates the efficiency of our heuristic checkpoint-placement strategy and we can quantify the costs associated with finding an optimal solution with \ilp and \DP.
To this end, we generate 100 distributions for varying numbers of steps and place checkpoints using all three methods (refer to \prettyref{tab:ilp}).

These experiments were conducted on an AMD Ryzen 7 Pro 5850U CPU (16 HW threads, 48\,GiB DRAM) and utilized Gurobi 10.0.1 to solve the ILP instance.
For the \genetic algorithm, we set a time limit of 10 seconds, and we recorded the moment at which the last improvement occurred—that is, when the solution initially stabilized at the final result.
In contrast, we restricted \ilp and \DP to run approximately 3 minutes; we employed Gurobi's standard options.
For more than 3000 steps, \ilp ran into the time limit and we cannot report run-time numbers.

A key observation, as evidenced by \prettyref{tab:ilp}, is that the \genetic algorithm arrived at the optimal selection, as discovered by \ilp or \DP, for \drefrel[percent of=/random/ilp/count]{/random/ilp/count_opt} percent of \change{all generated} distributions.
Only for \dref{/random/nonopt/count} distributions, all having 150\,000 steps, \genetic did not converge on the optimal solution.
For these three, however, the geometric mean of the over-approximation is only \drefcalc{d(/random/nonopt/over_gmean)*100-100} percent.

Further, we note that solving \ilp with Gurobi, which is a general-purpose solver not specifically target at our problem, scales worst.
With \DP, we can scale to 50 times larger problems within the same time limit.
In contrast, \genetic scales best for these distributions and converges to a final solution well within the time limit.
Hence, we deduce that the \genetic algorithm is well-suited to solve the checkpoint-placement problem, as it demonstrates both speed and efficiency in yielding favorable results.

\subsection{Real-World Distributions}
\label{sec:real-world}

Next, we want to explore the benefits our selecting methodology when applied to real-world \ac{FI} distributions that we derive from traces of the MiBench.
As these distributions have a high number of steps (up to 872 million), we only compare \uniform and \genetic as it is unrealistic to solve the resulting ILP instance or to execute~\DP.
For these experiments, we used an Ampere Altra machine with 80 aarch64 cores and 256\,GiB of DRAM.

\paragraph{Fault Model}
To explore different levels of \enquote{non-uniformity}, we chose a fault model that allows us to scale this metric while still being connected to a real hardware implementation:
With our fault model, all main-memory cells are uniformly vulnerable to single-event upsets (bit flips), while the caches are more robust against faults (due to being SRAM).
Since the cache \enquote{filters} the CPU's memory accesses, only some accesses actually access the memory, from where the system incorporates and, thereby, activates faults.
To exploit this bursty fault-activation pattern, a \ac{FI} planner would employ def-use pruning~\cite{sieh:97:ftcs} and inject faults at cache-miss time.

\begin{figure*}
  \centering
  \includegraphics[width=\textwidth]{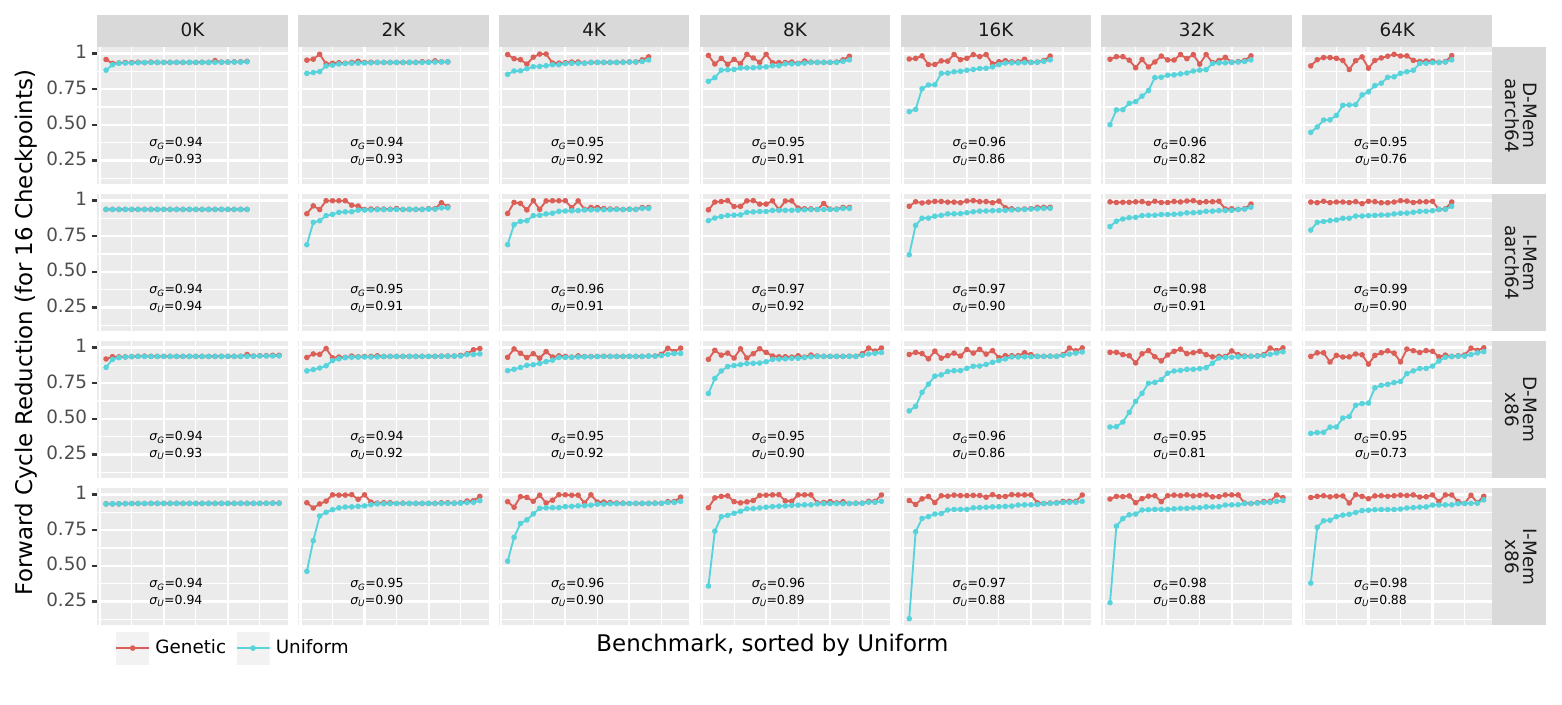}
  \caption{Forward-cycle reduction for different cache sizes and 16 checkpoints.
       The x-axis is sorted by the forward-cycle reductions for \uniform.
       $\sigma_G$ and $\sigma_U$ refer to the average reduction for \genetic and \uniform respectively.}\label{fig:by-caches}
\end{figure*}

\paragraph{MiBench Traces}
With the \texttt{valgrind} tool, we execute the MiBench benchmarks as Linux programs (aarch64, x86-64), and collect the memory-stage and the instruction-fetch accesses that happen after invoking \texttt{main()}.
We choose x86-64 as a representative for CISC architectures and aarch64 as representative for load-store RISC architectures.
With the obtained access traces, we use the \texttt{pycachesim} cache simulator~\cite{hammer:pycachesim} to derive the cache-miss distribution for different instruction- and data-cache setups: we simulate four-way associative caches with six sizes that range from 2\,KiB to 64\,KiB, which reflects the cache hierarchies of the safety-relevant Arm M7 processor family.

For x86 the benchmarks ispell, sphinx and rsynth from the office branch and the tiff's and mad from the consumer branch are not compilable.
For aarch64, the pgp\_\{d,e\} benchmarks is buggy while valgrind crashes for ghostscript and rijndael\_\{d,e\} due to a known bug.
With the two architectures (aarch64/x86-64), \dref{/benchmarks/aarch64}/\dref{/benchmarks/x86} benchmarks, two memory-access paths, and 7 cache sizes (including no cache), we end up with \drefcalc{(d(/benchmarks/aarch64) + d(/benchmarks/x86)) * 2 * 7} \ac{FI} distributions.
We use these distributions as~$D(t)$ and apply \uniform and \genetic, which we execute for 10 seconds.

\drefassert{d(/distributions/total)==((d(/benchmarks/aarch64) + d(/benchmarks/x86)) * 2 * 7)}

\drefkeys{P/.style={precision=2,assume math mode=true,percent}}
\csdef{MemString:True}{I-Mem}
\csdef{MemString:False}{D-Mem}
\def\MemString#1{\csuse{MemString:\drefvalueof{#1}}}
\def\Identify#1{\dref*{#1/benchmark}, \MemString{#1/icache}, \dref{#1/cachesize}K}

\paragraph{Fixed Number of Checkpoints}
To show that \genetic produces consistent results, we evaluate the selection strategies under different parameters: the cache size, difference in architecture and the number of checkpoints.
First, we quantify the influence of the cache size for different architectures (aarch64, x86-64) and memories (instruction, data).
While larger caches result in less uniform distributions, they have an especially high impact in the instruction-memory accesses as loops result in a high locality.

In \prettyref{fig:by-caches}, we show the reductions for selecting a fixed amount of 16 checkpoints over different cache sizes and architectures.
While \uniform is able to result in large reductions for many benchmarks, we also see that its results significantly deteriorate for large cache sizes.
Over the shown matrix, we see that \genetic ($\sigma_G \in [\dref[P]{/by-caches/genetic/min/value},\dref[P]{/by-caches/genetic/max/value}]$) consistently outperforms \uniform ($\sigma_U \in [\dref[P]{/by-caches/uniform/min/value},\dref[P]{/by-caches/uniform/max/value}]$).

When looking at individual benchmarks, \genetic achieves at least a reduction by \dref[P]{/by-caches/benchmark/genetic/min/value} percent (for \Identify{/by-caches/benchmark/genetic/min}), while \uniform even resulted in \emph{no improvement} for one benchmark (\Identify{/by-caches/benchmark/uniform/min}) as all injections were planned before the first uniform checkpoint.
\change{In the best case}, \genetic even achieves \dref[P,precision=3]{/by-caches/benchmark/genetic/max/value} percent (for \Identify{/by-caches/benchmark/genetic/max}) savings.
Regarding the architecture, we see no significant difference between aarch64 and x86, which brings us to the conclusion that our findings are also generalizable to other architectures.
\drefassert{d(/by-caches/benchmark/uniform/min/cardinality) == 1}

\begin{figure}
       \centering
	\includegraphics[width=\linewidth]{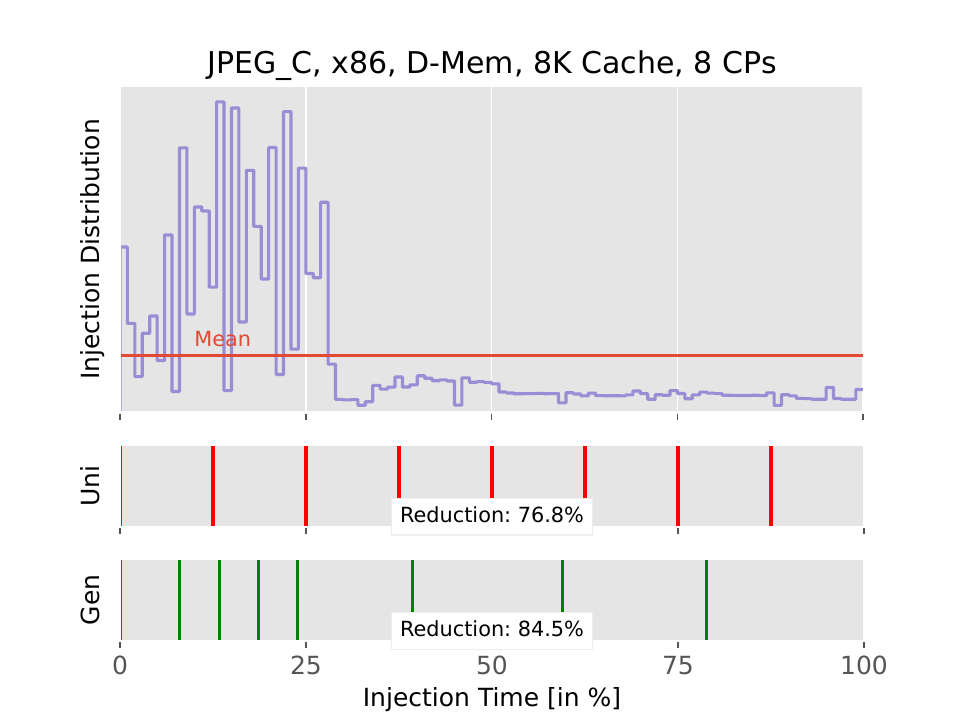}
	\caption{Distribution and checkpoints for JPEG compression. After the cache has warmed ($t\approx 30$), the data memory is only rarely read, leading to a skewed injection distribution.}
	\label{fig:distr}
\end{figure}

Qualitatively, we could identify three patterns:
(1) when everything fits into the cache, cache misses only occur in the warm-up period and \genetic correctly sets the checkpoints in the warm-up period, while \uniform distributes them blindly over the whole program run. We could observe this for the D-Mem of all benchmarks with a small input (\eg bitcount, ADPCMs and stringsearch).
(2) for benchmarks with a high cache pressure, cache misses occur regularly, and the injection instructions become more uniformly distributed. For these benchmarks (\eg PATRICIA with its >270\,KiB input size), the advantage of \genetic disappears.
(3) for benchmarks with an irregular cache-miss distribution, \uniform often places checkpoints in periods of low cache-miss rates, and the effect of the checkpoint is not optimally utilized.
For example, in \prettyref{fig:distr}, \uniform disadvantageously places checkpoints in a period with nearly no misses, while \genetic uses those in initial cache warming phase, leading to a \drefcalc{84.5-76.8} percentage-point improvement.

\begin{figure}
	\centering
        \includegraphics[width=\linewidth]{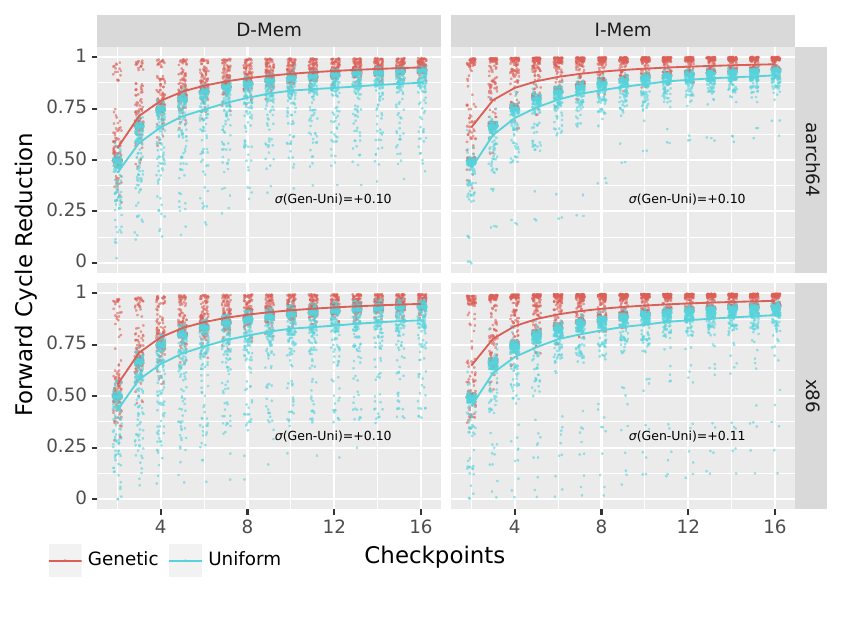}
	\caption{Varying Number of Checkpoints. Effect of the checkpoint count on the forward-cycle reduction over all cache sizes and benchmarks. Horizontal jitter to reduce overprinting.
        }
	\label{fig:cp_effect}
\end{figure}

\paragraph{Varying Number of Checkpoints}
Next, we are interested how both strategies perform when we scale the number of checkpoints between 2 and 16.
The results are shown in \prettyref{fig:cp_effect}, \change{where we plot the achieved reduction per benchmark as two points (one red and one blue).
The lines mark the average reduction per checkpoint count and placement strategy, while higher is better.
Further, we tile the results along the CPU architecture and cache-type axis to determine if those dimensions have an significant impact on the achieved savings.}

First, we can see that increasing the number of checkpoints has a diminishing effect for both strategies and the 16th checkpoint has a far smaller effect than the third one.
However, on average, \genetic has a consistent advantage, regardless of the memory kind or the architecture, which \uniform cannot close, even with 16 checkpoints.
At worst, and averaged over all benchmarks, \genetic requires \dref{/by-CPs/outperfom/CPs/max/CPs} checkpoints for \dref*{/by-CPs/outperfom/CPs/max/arch} \MemString{/by-CPs/outperfom/CPs/max/icache} to achieve the same reduction as 16 uniform checkpoints.
For \dref*{/by-CPs/outperfom/CPs/min/arch} \MemString{/by-CPs/outperfom/CPs/min/icache}, we even require only \dref{/by-CPs/outperfom/CPs/min/CPs} checkpoints to achieve an average reduction of \dref[P]{/by-CPs/outperfom/CPs/min/genetic} percent (uniform: \dref[P]{/by-CPs/outperfom/CPs/min/uniform}\,\%).

\begin{figure}[t]
	\centering
        \includegraphics[width=\linewidth]{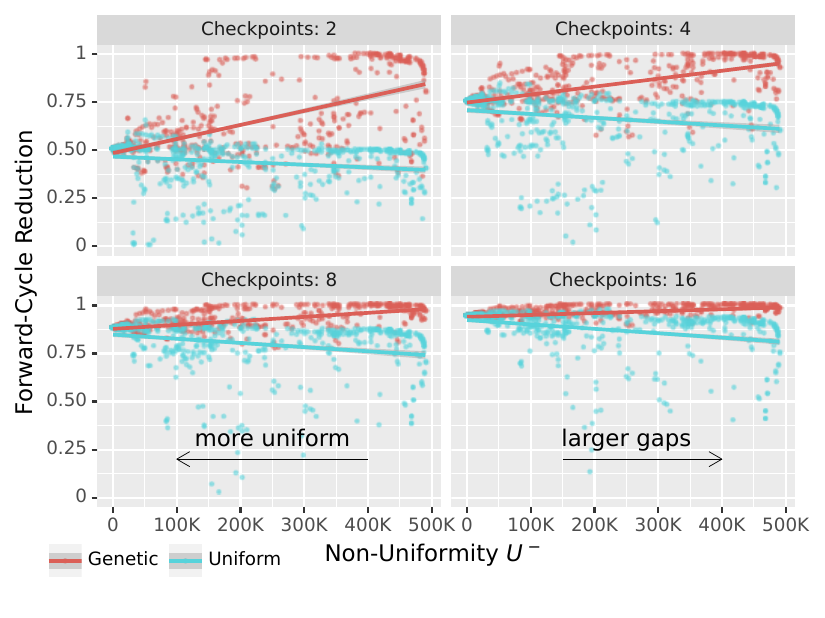}
	\caption{Sensitivity with Respect to the Non-Uniformity of the Fault Distribution. (N=\dref[assume math mode=true]{/distributions/total})}
	\label{fig:sensitivity}
 \end{figure}

 \paragraph{Sensitivity with Respect to Non-Uniformity}
Next, we investigate the influence of the "non-uniformity" of our real-world  distributions to further substantiate our conjecture that \genetic has superior performance over \uniform for less uniform inputs.
In \prettyref{fig:sensitivity}, \change{we plot the achieved} forward cycle reductions per benchmark against $U^-$.
Again, each benchmark appears as two points (red for \genetic and blue for \uniform) per tile. To highlight the trend for more non-uniform distributions, we plot a linear regression through both result sets.
We show results for four checkpoint counts (2, 4, 8, 16) and higher is better.

First, we see that real-world \ac{FI} distributions that stem from our fault model have an even higher \ac{WFFT} score than our synthetic benchmarks.
Making the checkpoint selection problem even more important for real-world \ac{FI} campaigns.

Further, while \uniform's performance deteriorates for less uniform distributions, \genetic's performance even exhibits an improvement, particularly when the number of available checkpoints is small.
This can be attributed to the fact that a checkpoint placed immediately prior to a significant peak yields a greater reduction than one positioned before an extended, shallow hill.

\section{Discussion}
\label{sec:discussion}
We evaluated the proposed checkpoint-selection algorithm on the MiBench benchmark suite for both aarch64 and x86, with varying numbers of checkpoints and cache sizes.
In addition, we demonstrated that \genetic is able to the achieve (almost) the same optimal results as \ilp and \DP while it scales better when confronted with larger problems.

Our evaluation reveals a greater efficiency in our checkpoint-placement methods for more irregular distributions.
These gains are attributed to our adherence to the actual distribution rather than the blind, uniform placement of checkpoints.
For a fault model that provoked non-uniform distributions, our method correctly pinpoints and leverages areas of high importance such as the cache warm-up period, characterized by a high density of injection sites.
Our approach, irrespective of architecture and memory type, consistently outperforms \uniform in selecting superior checkpoints and reducing forward-phase cycles.

The premise of this paper rests on the assumption of a fixed number of checkpoints, possibly constrained by the \ac{FI} platform.
Yet in situations with an unrestricted number of checkpoints, they are not without cost.
Firstly, checkpoints require memory and storage, potentially significant if the entire DRAM state is captured.
Secondly, the process of creating, storing, and distributing checkpoints to the fault injector consumes time, effectively reducing the net savings; more checkpoints may paradoxically lead to fewer savings.
Thirdly, our evaluation demonstrated diminishing returns from additional checkpoints.
In contrast, improving the placement of existing checkpoints enhances their effectiveness with marginal cost increase.
For instance, even for the worst outcome, \genetic with six checkpoints achieves a greater reduction than \uniform with 16 checkpoints.
However, determining the Pareto-optimal number of checkpoints for distribution-aware placement remains an area for future research.

Additionally, the overhead incurred by determining checkpoint distribution is minor compared to the cycles saved.
In our experiments, we limited \genetic's runtime to 10 seconds, although it often converged sooner (see \prettyref{tab:ilp}).
The total runtime of a complete systematic \ac{FI} campaign, while indeed dependent on the specific \ac{PUT}, typically spans several hours or even days.
Consequently, investing an additional ten seconds to optimize checkpoint placement using \genetic is always justifiable.
Given these considerations, we argue that an optimized checkpoint selection ought to become the norm for any \ac{FI} campaign that has, either by a constraint or by a design decision, a fixed number of checkpoints.

\section{Related Work}
\label{sec:related-work}

When we look into literature, \ac{FI} tools and, when reported, the way checkpoints are placed in evaluations can be divided into three categories.
The first common approach is to use them to skip the startup sequence of the simulator, which is often longer than the loading time for a checkpoint.
This is done in both \acl{FI} tools GemFI~\cite{parasyris:14:dsn} and MEFISTO~\cite{jenn:94:ftcs}.
In addition, several works report utilizing checkpoints in this way to accelerate their evaluation~\cite{amarnath:18:issrew,mahmoud:19:asplos}.
The next approach is to use more than a single checkpoint.
For example, GangES~\cite{hari:14:isca} saves checkpoints periodically during recording the golden run, which results in a uniform distribution.
Several other works~\cite{berrojo:02:vlsi,berrojo:02:date,tuzov:16:ladc,rosa:15:dfts} distribute them uniformly.
However, in all this works the distribution of checkpoints is not a focus, and thus they do not report on the effect of using checkpoints.
Some \ac{FI} tools like FAIL*~\cite{schirmeier:15:edcc} leave the decision, where to place checkpoints and how many to the user.

Very few studies attempt to quantify the impact of checkpoints on \ac{FI}-campaign run time.
\textcite{ruano:08:isie} positioned a single checkpoint at three-quarters of the total runtime and "almost at the end".
Their analysis found that the later checkpoint led to more significant runtime savings and they concluded that a detailed examination of checkpoint selection is necessary.
Parotta \emph{et al.}~\cite{parrotta:00:ioltw} \emph{uniformly} place checkpoints to accelerate hardware-assisted \ac{FI}-campaigns.
They find that beyond a certain number of checkpoints---in their case, 10 checkpoints---, savings become diminishing; a result that aligns with our own.
\textcite{schirmeier:14:ets} propose smart-hopping, an improved forwarding mechanism based on hardware breakpoints, to speed up the forwarding phase for hardware-assisted fault injection.
Although they briefly explore checkpoint placement, their placement method results in the \emph{uniform} distribution if used without smart-hopping and hardware support.
In contrast, we provide a fundamental study of checkpoint selection that is hardware and \ac{FI}-mechanism independent.

While not being our focus, the efficient storage and retrieval of checkpoints is another important topic~\cite{bautista:10:ccgrid}.

\section{Conclusion}
\label{sec:conclusion}
One cost factor of comprehensive \ac{FI} campaigns is the forwarding phase, which is the time required to bring the \acf{PUT} into the fault-free state at injection time.
The common technique to speed up this process are checkpoints of the fault-free system state at fixed points in time.
In this paper, we show that the placement of checkpoints has a significant influence on the required forwarding cost, especially if the planned faults are non-uniformly distributed in time.
For this, we discuss the checkpoint-selection problem in general, reduce it to the problem of finding the maximum-weight reward path in \acp{DAG}, and propose three distinct methods; two of them provide the optimal solution, while the third is a heuristic based on genetic algorithms.

We compared the proposed methods with synthetic benchmarks and applied our genetic algorithm on the MiBench benchmark suite on both aarch64 and x86, with varying amounts of checkpoints and cache size reflecting those of the M7 processor family.
This evaluation parameters resulted in a total of \drefcalc{(d(/benchmarks/aarch64) + d(/benchmarks/x86)) * 2 * 7} \ac{FI} distributions.
Our approach consistently performs better than a time-uniform checkpoint selection regardless of the underlying architecture and equally for data- and instruction-fetch accesses.
Overall, with 16 checkpoints, we are able to consistently reduce the forward-phase cycles, with reductions of at least \dref[PA,percent,precision=0]{/by-caches/benchmark/genetic/min/value} percent and up to \dref[PA,percent,precision=3]{/by-caches/benchmark/genetic/max/value} percent.

\section*{Acknowledgements}
\noindent
We thank the anonymous reviewers (in advance) for their valuable feedback and dedicated efforts in helping us improve this paper.
This work was funded by the \emph{Deutsche Forschungsgemeinschaft (DFG, German Research Foundation)} – 468988364, 501887536.

\begin{mdframed}[linecolor=safegreen,backgroundcolor=safegreen!30,linewidth=3pt]
  The source code and fault distributions used for the evaluation are available as an data artifact~\cite{dietrich:23:iccad-artifact}.
\end{mdframed}

\printbibliography
\end{document}